\documentclass[a4paper,twoside]{article}
\newcommand{\affil}[1]{$^{\rm #1}$}
\usepackage[authoryear]{natbib}
\bibpunct{(}{)}{;}{a}{}{,}
\usepackage{graphicx}
\date{} %Please leave the date blank
\title{\large\bf\flushleft A concise reference to (projected) S\'ersic $R^{1/n}$ quantities, including Concentration, Profile Slopes, Petrosian indices, and Kron Magnitudes}
\author{\parbox{\textwidth}{\flushleft
\vspace{-0.5cm}
{\it 
Alister W.\ Graham\affil{A,}\affil{B} and Simon P.\ Driver\affil{A}
}\\
\vspace{0.4cm}
{\small \affil{A}\,Mount Stromlo and Siding Spring Observatories, Australian National University, Private Bag, Weston Creek PO, ACT 2611, Australia.}\\
{\small \affil{B}\,Email: Graham@mso.anu.edu.au}}}
\begin{document}
\twocolumn[
\begin{changemargin}{.8cm}{.5cm}
\begin{minipage}{.9\textwidth}
\vspace{-1cm}
\maketitle
%
%
%%%%%%%%%%%%%     ABSTRACT    %%%%%%%%%%%%%
%Abstract of no more than 200 words here.
\small{\bf 
Abstract: Given the growing use of S\'ersic's (1963, 1968) $R^{1/n}$
model for describing the stellar distributions in galaxies, and the
lack of any single reference that provides the various associated
mathematical expressions, we have endeavoured to compile such a
resource here.  We present the standard intensity profile, and its
various guises such as the luminosity, surface--brightness, and
aperture--magnitude profile.  Expressions to transform the effective
surface brightness into the mean effective and central surface
brightness are also given, as is the expression to transform between
effective radii and exponential scale--lengths.  We additionally
provide expressions for deriving the `concentration' of an $R^{1/n}$
profile, and two useful equations for the logarithmic slope of the
light--profile are given.
Petrosian radii and fluxes are also derived for a range of S\'ersic
profiles and compared with the effective radii and total flux.
Similarly, expressions to obtain Kron radii and fluxes are presented
as a function of the S\'ersic index $n$ and the number of effective
radii sampled.  Illustrative figures are provided throughout.
Finally, the core--S\'ersic model, consisting of an inner
power--law and an outer--S\'ersic function, is presented.
}

%%%%%%%%%%%%%     KEYWORDS    %%%%%%%%%%%%%
\medskip{\bf Keywords:}   galaxies: structure,  galaxies: fundamental parameters, methods: analytical,  methods: data analysis
% Please write all keywords in lower case. PASA uses the
% standard list of subject headings adopted by The Astrophysical Journal
% and available from http://www.journals.uchicago.edu/ApJ/keywords_text.html.
% Keywords are separated by em-dashes, i.e. ---

%%%%%%%%DO NOT EDIT%%%%%%%%%%%%
\medskip
\medskip
\end{minipage}
\end{changemargin}
]
\small
%%%%%%%%EDIT FROM HERE%%%%%%%%%%%%

\section{Introduction}

Working with the 30-inch Reynolds telescope\footnote{The Reynolds
telescope was sadly destroyed in the 2003 Canberra bush fires.} at
what was then Australia's Commonwealth Observatory, and today known as
Mount Stromlo Observatory, G\'erard de Vaucouleurs published in 1956 the
most extensive southern galaxy Atlas of the day.  In the following
year, Jos\'e Luis S\'ersic commenced work at the 1.54-m telescope at
the Astrophysical Station at Bosque Alegre in Argentina.  His studies
from 1957--1966 culminated in his 1968 southern--hemisphere galaxy
Atlas `Galaxias Australes'.  It too has proven an invaluable
contribution to our understanding of galaxies, evidenced by its status
as a top 1000 cited astronomy publication.

In the Introduction of S\'ersic's Atlas, it not only states the merits
for a visual representation of galaxies, but, like de Vaucouleurs', it
stresses the necessity to go beyond this and obtain quantitative
measures of the light distribution.  This was not mere rhetoric as
his Atlas consists of two parts, one pictorial in nature and the
latter quantitative.
It is apparent that his generalisation of de Vaucouleurs' (1948, 1959)
$R^{1/4}$ model to an $R^{1/n}$ model was not merely something he
mentioned in passing, but something which he felt {\it should} be
done.  Indeed, S\'ersic fitted the $R^{1/n}$ model to every
(sufficiently large) galaxy in his Atlas.  He derived expressions to
compute total (extrapolated) galaxy magnitudes, provided tables of
assorted structural parameters associated with the $R^{1/n}$ model,
and showed how they correlated with galaxy morphological type (his
Figure 3) and galaxy concentration (his Figure 4, page 145).
S\'ersic (1963) even provides a
prescription to correct the $R^{1/n}$ model parameters for 
Gaussian seeing due to atmospheric and instrumental dispersion. 

It is, however, of interest to note that S\'ersic's conviction lay in
the observation that different galaxies possessed differing degrees of
an $R^{1/4}$ bulge and an $R^{1/1}$ disk component. This mixture of bulge
and disk components produces a combined surface brightness profile 
with an intermediary form, hence the $R^{1/n}$ model.

Today, usually when the required resolution is lacking to properly
decompose an image into its separate bulge and disk components,
galaxies are modelled with a single $R^{1/n}$ profile, just as
S\'ersic proposed (e.g., Blanton et al.\ 2003).  
% 
% This paper incorrectly binned data equally in $n$ rather than `log(n)'
% and thus under-populated the higher-n bins.  It is log(n), not n, 
% that correlates linearly with magnitude. 
% 
While such an approach certainly has its merits,
we now know that dynamically hot stellar systems themselves posses a
range of profile shapes that are well described with the $R^{1/n}$
model (e.g., Graham \& Guzm\'an 2003, and references therein).  
Detailed studies of well resolved lenticular and disk galaxies 
are routinely fitted with the combination 
of an exponential--disk plus an $R^{1/n}$--bulge
(e.g., Andredakis, Peletier, \& Balcells 1995; Seigar \& James 1998;
Iodice, D'Onofrio, \& Capaccioli 1997, 1999; 
Khosroshahi, Wadadekar, \& Kembhavi 2000; 
D'Onofrio 2001; Graham 2001a; M\"ollenhoff \& Heidt 2001). 
In either case, since the work of Capaccioli in the late 1980s and in
particular Caon, Capaccioli, \& D'Onofrio (1993) and D'Onofrio,
Capaccioli, \& Caon (1994), the past decade has seen an explosion in
the application of the $R^{1/n}$ model (e.g., Cellone, Forte, \&
Geisler 1994; Vennik \& Richter 1994; Young \& Currie 1994, 1995;
Graham et al.\ 1996; Karachentseva 1996, Vennik et al.\ 1996, to
mention just a few early papers), yet no single resource exists 
for the expressions and
quantities pertaining to the $R^{1/n}$ model.  Moreover, no one
reference provides more than a few of the relevant equations, and many
textbooks still only refer to the $R^{1/4}$ model.

This (largely review) article intends to provide a compendium of
equations, numbers, and figures for ease of reference.  The derivation
of these also provide useful exercises for students.  Where
appropriate, we have endeavoured to cite the first, or at least a
useful early, reference to any given equation.
To the best of our knowledge, Figures~(\ref{figPet1}) through (\ref{figKronL2}), 
describing Petrosian indices and Kron magnitudes, have never been seen before. 
% We also include (possibly new) expressions that show how Petrosian radii
% and fluxes, and also Kron radii and fluxes, relate to S\'ersic
% profiles.
%
A brief reference to where readers can find deprojected expressions, and 
how to deal with practical issues such as seeing, is
given at the end.  No attempt has been made here to show the numerous
scientific advances engendered via application of the $R^{1/n}$ model.

\section{S\'ersic related quantities}

\subsection{The S\'ersic profile \label{SectOne}} 

S\'ersic's (1963; 1968) $R^{1/n}$ model is most commonly expressed as an 
intensity profile, such that 
\begin{equation}
I(R)=I_{\rm e}\exp\left\{ -b_n\left[\left( \frac{R}{R_{\rm e}}\right) ^{1/n} -1\right]\right\},
\label{One} 
\end{equation}
where $I_{\rm e}$ is the intensity at the effective radius $R_{\rm e}$
that encloses half of the total light from the model (Caon 
et al.\ 1993; see also Ciotti 1991, his Equation 1).
The constant $b_n$ is defined in terms of the third and final
parameter $n$ which describes the `shape' of the light--profile, and
is given below.\footnote{It is common for researchers studying dwarf
galaxies to replace the exponent $1/n$ with $n$.  In this case, de
Vaucouleurs' model would have $n=0.25$, rather than 4.}

One can integrate Equation~(\ref{One}) over a projected area 
$A=\pi R^2$ to obtain the luminosity, $L$, interior to any 
radius R.  This is 
simply a matter of solving the integral\footnote{Obviously if one is using 
a major-- or minor--axis profile, rather than the geometric mean 
($R=\sqrt{ab}$) profile, an ellipticity term will be required. 
This is trivial to add and for the sake of simplicity won't be 
included here.  The issue of ellipticity gradients is more difficult, 
and interested readers should look at Ferrari et al.\ (2004).} 
\begin{eqnarray}
L(<R) = \int_0^R I(R^{\prime})2\pi R^{\prime} {\rm d}R^{\prime}, \nonumber
\end{eqnarray}
which yields, after substituting in $x=b_n(R/R_{\rm e})^{1/n}$, 
\begin{equation}
L(<R)=I_{e}R_{\rm e}^{2}2\pi n\frac{{\rm e}^{b_n}}{(b_n)^{2n}}\gamma (2n,x), 
\label{Lum}
\end{equation}
where $\gamma $(2n,x) is the incomplete gamma function defined by
\begin{equation}
\gamma (2n,x)=\int ^{x}_{0} {\rm e}^{-t}t^{2n-1}{\rm d}t.
\label{gamFunc}
\end{equation}
Replacing $\gamma (2n,x)$ with $\Gamma (2n)$ in Equation~(\ref{Lum})
gives one the value of $L_{tot}$ (Ciotti 1991).

Thus, the value of $b_n$ which we saw in Equation~(\ref{One}) is such 
that 
\begin{equation}
\Gamma (2n)=2\gamma (2n,b_n), 
\label{butt}
\end{equation} 
where $\Gamma $ is the (complete) gamma function (Ciotti 1991). 
Common values of $b_n$ are $b_4 = 7.669$ and $b_1 = 1.678$.
In passing we note a useful property of the $\Gamma$ function, which
is, $\Gamma (2n) = (2n-1)!$.

Analytical expressions which approximate the
value of $b_n$ have been developed.  Capaccioli (1989) provided one of
the first such approximations such that $b_n = 1.9992n-0.3271$, for
$0.5 < n < 10$ (see also Prugniel \& Simien 1997, their equation A3a).  
Ciotti \& Bertin (1999) showed $b_n \rightarrow
2n-1/3$ for large values of $n$, and in practice this provides a
better fit for values of $n$ greater than about 8 
(see Graham 2001a, his Figure 2). Ciotti \&
Bertin (1999) also provided an asymptotic expansion which, for values
of $n>0.36$, is accurate to better than 10$^{-4}$ and the
approximation of choice.  For smaller values of $n$, MacArthur,
Courteau, \& Holtzman (2003) provide a fourth order polynomial which
is accurate to better than two parts in 10$^3$ (see their Figure 23).
However, the exact value of $b_n$ in Equation~(\ref{butt}) can be solved 
numerically, and fast codes exist to do this, 

For an exponential ($n=1$) profile, 99.1\% of the flux resides within
the inner 4 $R_{\rm e}$ (90.8\% within the inner 4 scale--lengths) and
99.8\% of the flux resides within the inner 5 $R_{\rm e}$ (96.0\%
within the inner 5 scale--lengths).
For an $n=4$ profile, 84.7\% of the flux resides with the inner 4
$R_{\rm e}$ and 88.4\% within the inner 5 $R_{\rm e}$.

\begin{figure}[ht]
\begin{center}
\includegraphics[scale=0.5, angle=270]{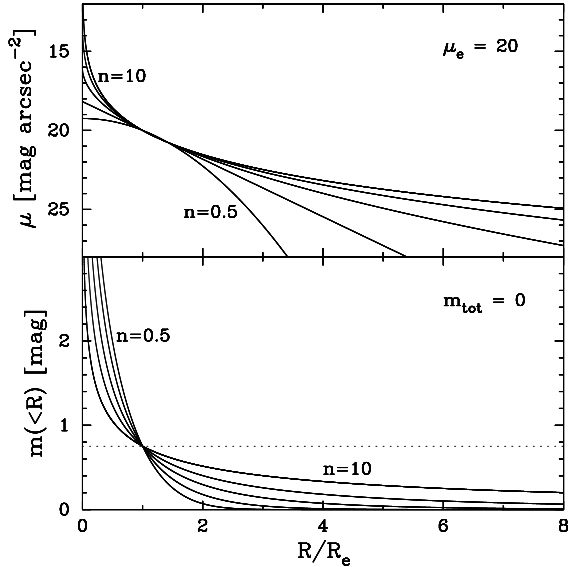}
\caption{
Top panel: S\'ersic surface brightness profiles (Equation~\ref{EqSB}) 
for $n$=0.5, 1, 2, 4, and 10. The profiles have been normalised
at $\mu_{\rm e}=20$ mag arcsec$^{-2}$.
Bottom panel: S\'ersic aperture magnitude profiles (Equation~\ref{Eqmag}),
normalised such that the total magnitude equals zero.   The dotted 
line is offset by 0.75 mag (a factor of 2 in flux) from the total 
magnitude. 
}
\label{figSersic}
\end{center}
\end{figure}

Multiplying the negative logarithm of the luminosity profile 
(Equation~\ref{Lum}) by 2.5 gives the enclosed magnitude profile, 
known as the ``curve of growth'', 
\begin{equation}
m(<R) = \mu_{\rm e} - 5\log R_{\rm e} -2.5 \log \left[ 2\pi n\frac{{\rm e}^{b_n}}{(b_n)^{2n}}\gamma (2n,x) \right], 
\label{Eqmag}
\end{equation}
which asymptotes to the total apparent magnitude $m_{tot}$ as $R$
tends to infinity and, consequently, $\gamma (2n,x) \rightarrow \Gamma
(2n)$ (Figure~\ref{figSersic}).

Multiplying the negative logarithm of Equation~(\ref{One}) by 2.5
yields the surface brightness profile (Figure~\ref{figSersic}), as
used in Caon et al.\ (1993),
\begin{equation}
\mu (R)=\mu_{\rm e}+\frac{2.5b_n}{\ln(10)}\left[\left(R/R_{\rm e}\right)^{1/n}-1\right].
\label{EqSB}
\end{equation}

\subsubsection{Asymptotic behavior for large $n$}

For large values of $n$, the S\'ersic model tends to a power--law with 
slope equal to 5. 
Substituting ${\rm e}^t = z = R/R_{\rm e}$ into Equation~(\ref{One}), one has 
\begin{eqnarray}
I(z)  \sim \exp\left\{ -b_n\left[ {\rm e}^{t/n} -1 \right] \right\}. \nonumber 
\nonumber
\end{eqnarray}
Now, for large $n$, ${\rm e}^{t/n}$ is small, and so 
one can use e$^{t/n} \sim 1+t/n$. One can also use $b_n \sim 2n$ to give
\begin{eqnarray}
\ln[I(z)] \sim  -b_n\left[ t/n \right] \sim -2t \sim -2\ln(z). \nonumber
\end{eqnarray} 
Thus $\mu(z) = -2.5 \log[I(z)] \sim 5\log(z)$. 

\vspace{2mm}
For simplicity, the subscript `$n$' will be dropped from the term $b_n$ in 
what follows.

\subsection{Surface brightness, radial scale, and absolute magnitude}

From the value of $\mu_{\rm e}$, the `effective surface brightness' at
$R_{\rm e}$, and knowing the value of $n$, one can compute both the
central surface brightness $\mu_{\rm 0}$ and the average/mean surface
brightness $\langle\mu\rangle_{\rm e}$ within the effective radius. 

\begin{figure}[ht]
\begin{center}
\includegraphics[scale=0.42, angle=270]{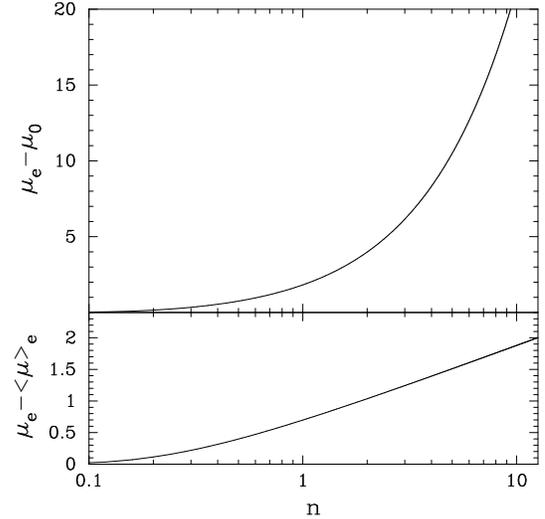}
\caption{
Top panel: Difference between the central surface 
brightness $\mu_{\rm 0}$ and the effective surface brightness $\mu_{\rm e}$
as a function of profile shape $n$.  Bottom panel: 
Difference between $\mu_{\rm e}$ and the mean effective surface brightness 
$\langle\mu\rangle_{\rm e}$ as a function of $n$.
}
\label{figMuDiff}
\end{center}
\end{figure}

At the centre of the profile one has, from Equation~(\ref{EqSB}), 
\begin{eqnarray}
\mu(R=0)\equiv \mu_{\rm 0} = \mu_{\rm e} - 2.5b/\ln (10), \\
\mu_{\rm 0} = \mu_{\rm e} - 1.822, n=1, \nonumber \\
\mu_{\rm 0} = \mu_{\rm e} - 8.327, n=4. \nonumber \\
\nonumber
\end{eqnarray}
The difference between $\mu_{\rm e}$ and $\mu_{\rm 0}$ is shown
in Figure~(\ref{figMuDiff}) as a function of the S\'ersic index $n$. 

The `mean effective surface brightness', often simply referred to as
the `mean surface brightness', is computed 
as follows.  The average intensity, $\langle I \rangle_{\rm e}$, 
within the effective radius is obtained by integrating the intensity 
over the area $A=\pi R_{\rm e}^2$ such that 
\begin{eqnarray}
\langle I \rangle_{\rm e}=\frac{\int I dA}{A}=\frac{I_{\rm e}{\rm e}^{b}\int _{0}^{R_{\rm e}}{\rm e}^{-b\left(R/R_{\rm e}\right)^{1/n}} 2\pi R {\rm d}R}{\pi R_{\rm e}^{2}}.  \nonumber 
\end{eqnarray}
Letting $x=b(R/R_{\rm e})^{1/n}$, {\rm one has} 
\begin{eqnarray}
\langle I \rangle_{\rm e}=I_{\rm e}f(n), \nonumber
\end{eqnarray}
where
\begin{eqnarray}
f(n)=\frac{2n {\rm e}^{b}}{b^{2n}}\int_{0}^{b}{\rm e}^{-x}x^{2n-1} {\rm d}x. \nonumber 
\end{eqnarray}
Now as $b$ was chosen such that $R_{\rm e}$ is the radius containing
half of the total light, one has 
\begin{equation}
f(n)=\frac{n {\rm e}^{b}}{b^{2n}}\int_{0}^{\infty}{\rm e}^{-x}x^{2n-1} {\rm d}x=\frac{n {\rm e}^{b}}{b^{2n}}\Gamma (2n).
\end{equation}
Thus, 
\begin{eqnarray}
\langle\mu\rangle_{\rm e} = \mu_{\rm e} -2.5\log [f(n)], \label{EqMu} \\
\langle\mu\rangle_{\rm e} = \mu_{\rm e} - 0.699, n=1, \nonumber \\
\langle\mu\rangle_{\rm e} = \mu_{\rm e} - 1.393, n=4. \nonumber 
\end{eqnarray}
The difference between $\mu_{\rm e}$ and $\langle\mu\rangle_{\rm e}$
is shown in Figure~(\ref{figMuDiff}) as a function of the S\'ersic index
$n$ (Caon et al.\ 1994; Graham \& Colless 1997).

Substituting equation~\ref{EqMu} into Equation~\ref{Eqmag}, 
one has, at $R=R_{\rm e}$, 
\begin{equation}
m(<R_{\rm e}) = \langle\mu\rangle_{\rm e} - 2.5\log(\pi R_{\rm e}^2),
\end{equation}
and thus\footnote{Using empirical measurements within some suitably 
large aperture, one has from simple geometry that 
$\langle\mu\rangle_{1/2} = m_{\rm tot,ap} + 2.5\log(2\pi R_{1/2}^2)$.
Expressions to correct these approximate values --- due to the missed
flux outside of one's chosen aperture --- are given in Graham et al.\ (2005).}
\begin{equation} 
m_{\rm tot} = \langle\mu\rangle_{\rm e} - 2.5\log(2\pi R_{\rm e}^2), 
\end{equation} 
This expression can be rewritten in terms 
of the absolute magnitude, $M_{\rm tot}$, the effective radius
in kpc, $R_{\rm e, kpc}$, and the {\it absolute} effective
surface brightness, $\langle\mu\rangle_{\rm e, abs}$, (i.e., the mean
effective surface brightness if the galaxy was at a distance of 10pc):
\begin{eqnarray} 
M_{\rm tot} = \langle\mu\rangle_{\rm e,abs} - 2.5\log(2\pi R_{\rm e,kpc}^2) \nonumber \\
- 2.5\log\left[ \left( \frac{360\times 60\times 60}{2\pi \times 0.01} \right)^2 \right], \nonumber \\
M_{\rm tot} = \langle\mu\rangle_{\rm e,abs} - 2.5\log(2\pi R_{\rm e,kpc}^2) - 36.57 
\end{eqnarray}
The apparent and absolute mean effective surface brightnesses 
are related by the cosmological corrections:
\begin{equation}
\langle\mu\rangle_{\rm e,abs}=\langle\mu\rangle_{\rm e} -10\log(1+z)-E(z)-K(z), 
\end{equation}
where $z$, $E(z)$, and $K(z)$ are the redshift, evolutionary correction, 
and K-correction respectively (e.g., Driver et al.\ 1994 and references 
therein).

Another transformation arises from the use of scale--lengths $h$
rather than effective radii $R_{\rm e}$.  When the $R^{1/n}$ model is
written as 
\begin{equation}
I(R) = I_{\rm 0} \exp\left[ -(R/h)^{1/n} \right]
\label{oldie}
\end{equation}
% Ellis \& Perry (1979, their page 362, just before Eqn.\ 11a)
% Freeman 1970 refers to Sersic's Atlas for photometry purposes only
%              this occurs under Table 1. 
% 
(e.g., Ellis \& Perry 1979, their page 362; Davies et al.\ 1988), 
where $I_{\rm 0}=I(R=0)$, one has
\begin{eqnarray}
I_{\rm 0} = I_{\rm e} {\rm e}^b, \\ 
R_{\rm e}=b^n h, \\
R_{\rm e}=1.678h, n=1, \nonumber \\
R_{\rm e}=3466h, n=4. \nonumber 
% The PASA style 3 466h looks odd/confusing and has been avoided.
\end{eqnarray} 

It's straightforward to show that 
\begin{eqnarray}
\mu(R) = \mu_0 + \frac{2.5}{\ln(10)} \left( \frac{R}{h}\right) ^{1/n}, \\
\mu(R) = \mu_0 + 1.086(R/h), n=1, \nonumber
\end{eqnarray} 
and
\begin{equation}
L_{\rm tot} = \pi I_{\rm 0} h^2 \Gamma(2n+1). 
\end{equation} 
Given the small scale--lengths associated with the $n=4$ model, and
the practical uncertainties in deriving a galaxy's central brightness,
one can appreciate why Equation~(\ref{One}) is preferred over
Equation~(\ref{oldie}).

If one is modelling a two--component spiral galaxy, consisting of an
exponential disk and an $R^{1/n}$ bulge, then the bulge--to--disk
luminosity ratio is given by the expression
\begin{equation}
\frac{B}{D}=\frac{n\Gamma (2n) {\rm e}^{b}}{b^{2n}}
\left( \frac{R_{\rm e}^{2}}{h^2} \right)
\left( \frac{I_{\rm e}}{I_{\rm 0}} \right), 
\label{EqRad}
\end{equation}
where $h$ and $I_{\rm 0}$ are respectively the scale--length and
central intensity of the disk, and $R_{\rm e}, I_{\rm e}$, and $n$
describe the S\'ersic bulge profile.  Noting that $2n\Gamma(2n) =
\Gamma(2n+1) = (2n)!$, the above equation can be simplified for
integer values of $2n$.  For those who are curious, 
the first term on the right hand side of the
equality can be seen plotted as a function of $n$ in Graham (2001a).

\subsection{Concentration \label{SecCon}} 

For many years astronomers have had an interest in how centrally
concentrated a galaxy's stellar distribution is (e.g., Morgan 1958,
1959, 1962; Fraser 1972; de Vaucouleurs 1977).  
%
% Most recently, for dynamically hot systems (i.e., ellipticals or
% spiral galaxy bulges), the degree of concentration has been shown to
% be an accurate indicator of a system's supermassive black hole mass
% (Graham et al.\ 2001b).
%
S\'ersic (1968) used de Vaucouleurs (1956) somewhat subjective size
ratio between the main region of the galaxy and the apparent maximum
dimension of the galaxy as a measure of concentration.

\begin{figure}[ht]
\begin{center}
\includegraphics[scale=0.62, angle=270]{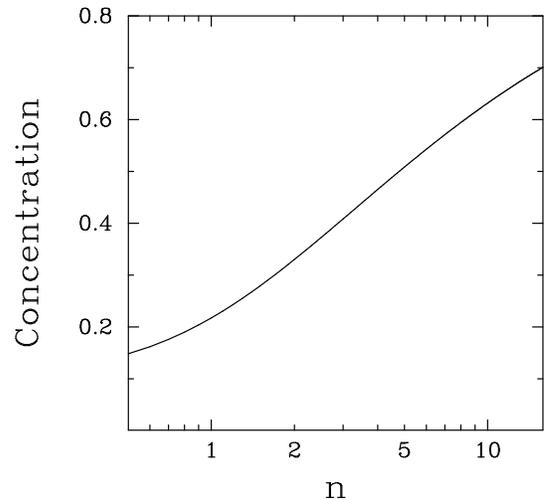}
\caption{The central concentration $C_{R_{\rm e}}(1/3)$,  
as defined by Trujillo et al.\ (2001c), is a monotonically increasing 
function of the S\'ersic index $n$.}
\label{figConc}
\end{center}
\end{figure}

Trujillo, Graham \& Caon (2001c) defined a useful, objective expression for 
concentration, such that, in pixelated form 
\begin{equation}
C_{R_{\rm e}}(\alpha)=\frac{\sum_{i,j\in E(\alpha R_{\rm e})}I_{ij}}{\sum_{i,j\in E(R_{\rm e})}I_{ij}}.
\label{Eqconc1}
\end{equation}
Here, $E(R_{\rm e})$ means the isophote which encloses half of the total light, 
and $E(\alpha R_{\rm e})$ is the
isophote at a radius $\alpha$ ($0<\alpha<1$) times $R_{\rm e}$.  
This is a flux ratio.  
For a S\'ersic profile which extends to infinity,
\begin{equation}
C_{R_{\rm e}}(\alpha)=\frac{\gamma(2n,b\alpha^{1/n})}{\gamma(2n,b)}.
\label{Eqconc2}
\end{equation}
This expression is a monotonically increasing function of $n$, and 
for $\alpha=1/3$ it's values are shown in Figure~(\ref{figConc}). 
An often unrealized point is that if every elliptical galaxy had an
$R^{1/4}$ profile then they would all have exactly the same degree of
concentration.  Observational errors aside, it is only because
elliptical galaxy light--profiles deviate from the $R^{1/4}$ model that
a range of concentrations exist.  This is true for all objective 
concentration indices in use today.

It should be noted that astronomers don't actually know where the
edges of elliptical galaxies occur; their light--profiles appear to
peeter--out into the background noise of the sky.  Due to the presence
of faint, high--redshift galaxies and scattered light, it is not
possible to determine the sky--background to an infinite degree of
accuracy.  From an analysis of such sky--background noise sources,
Dalcanton \& Bernstein (2000, see also Capaccioli \& de Vaucouleurs 1983) 
determined the limiting surface 
brightness threshold to be $\mu \sim 29.5$ $B$-mag arcsec$^{-2}$ and
$\mu \sim 29$ $R$-mag arcsec$^{-2}$.  Such depths are practically
never achieved and shallow exposures often fail to include a
significant fraction of an elliptical galaxy's light.  One would
therefore like to know how the concentration index may vary when
different galaxy radial extents are sampled but no effort is made to
account for the missed galaxy flux.  The resultant impact on
$C_{R_{\rm e}}$ and other popular concentration indices is addressed
in Graham, Trujillo, \& Caon (2001).

It was actually, in part, because of the unstable nature of the
popular concentration indices that Trujillo et al.\ (2001c) introduced
the notably more stable index given in Equations~(\ref{Eqconc1}) and
(\ref{Eqconc2}).   
The other reason was because the concentration index 
$C(\alpha) = \sum_{i,j\in E(\alpha)}I_{ij} / \sum_{i,j\in E(1)}I_{ij}$, 
where $E(\alpha)$ denotes some inner radius which is $\alpha$ ($0<\alpha<1$)
times the outermost radius which has been normalized to 1 
(Okamura, Kodaira, \& Watanabe 1984; Doi, Fukugita, \& Okamura 1993; 
Abraham et al.\ 1994),  
should tend to 1 for practically every profile that is sampled to
a large enough radius.  It is only because of measurement errors, 
undersampling, or the 
presence of truncated profiles such as the exponential disks in 
spiral galaxies, that this index deviates from a value of 1.

\subsection{Profile slopes}

\begin{figure}[ht]
\begin{center}
\includegraphics[scale=0.7, angle=270]{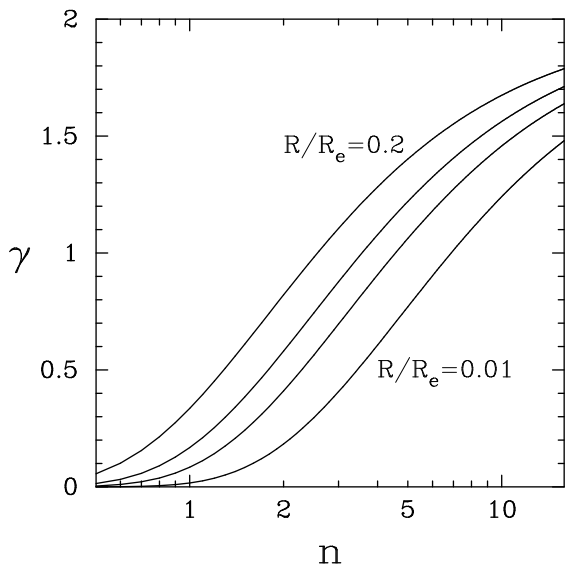}
\caption{
The slope of the S\'ersic profile $\gamma$ (Equation~\ref{Eqgam})
is shown as a function of profile shape $n$ for $R/R_{\rm e}$=0.01, 0.05,
0.1, and 0.2. 
}
\label{figGamma}
\end{center}
\end{figure}

\begin{figure}[ht]
\begin{center}
\includegraphics[scale=0.7, angle=270]{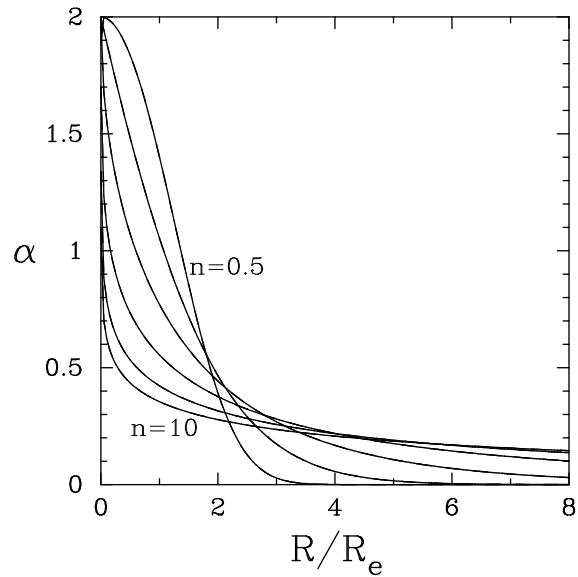}
\caption{
The slope $\alpha$
(Equation~\ref{alp}) is shown as a function of normalised radius $R/R_{\rm e}$
for $n$=0.5, 1, 2, 4, 7, and 10.
}
\label{figAlpha}
\end{center}
\end{figure}

Given {\sl HST's} ability to resolve the inner light--profiles of nearby
galaxies, the slope
$\gamma$ of a galaxy's nuclear (the inner few hundred parsec) stellar
distribution has become a quantity of interest. 
Defining\footnote{This $\gamma$ should not be confused with the incomplete
gamma function seen in Equation~(\ref{gamFunc}).}
\begin{equation}
\gamma(R) \equiv \frac{-{\rm d}[\log I(R)]}{{\rm d}\log R}, 
\end{equation} 
Rest et al.\ (2001, their Equation 8) used this to measure the nuclear
slopes of `core' and `power--law' galaxies. 
From Equation~(\ref{One}) one can obtain
\begin{equation}
\gamma(R,n) = (b/n)(R/R_{\rm e})^{1/n}.
\label{Eqgam} 
\end{equation} 
This is approximately $2(R/R_{\rm e})^{1/n}$ (see
section~\ref{SectOne}).  Thus, at constant $(R/R_{\rm e})$, $\gamma$ is a
monotonically increasing function of the S\'ersic index $n$ (Graham et
al.\ 2003b).  

It turns out Equation~(\ref{Eqgam}) is appropriate for the so--called
`power--law' galaxies which are now known to possess S\'ersic profiles
down to their resolution limit (Trujillo et al.\ 2004) and would be
better referred to as `S\'ersic' galaxies as they do not have
power--law profiles.  A modification is however required for the
luminous `core galaxies', and is described in Section~\ref{SecCore}.

Another logarithmic slope  of interest is that used by 
Gunn \& Oke (1975) and Hoessel (1980), 
% Hoessel mentions Petrosian (1976) on his page 503. 
and is defined as 
\begin{equation}
\alpha(R) \equiv \frac{{\rm d}[\ln L(R)]}{{\rm d}\ln R}. 
\label{EqnAlp}
\end{equation}
From Equation~(\ref{Lum}) one has
\begin{equation}
\alpha(x,n) = \frac{x}{nL(x)} \frac{{\rm d}[L(x)]}{{\rm d}x} 
            = \frac{{\rm e}^{-x}x^{2n}}{n\gamma (2n,x)}, 
\label{alp}
\end{equation}
where, as before, $x=b(R/R_{\rm e})^{1/n}$ (Graham et al.\ 
1996, their equation 8).

Figures~(\ref{figGamma}) and (\ref{figAlpha}) show how 
$\gamma(R)$ and $\alpha(R)$ vary with 
normalised radius $R/R_{\rm e}$ for a range of different profile shapes $n$.

\subsection{Petrosian index and magnitude} 

\begin{figure}[ht]
\begin{center}
\includegraphics[scale=0.4, angle=270]{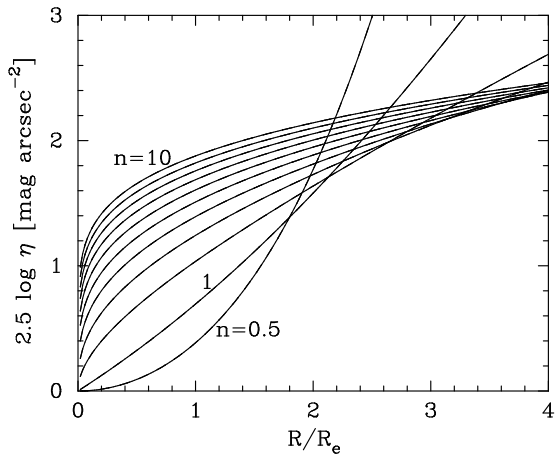}
\caption{
The logarithm of the Petrosian function $\eta(R)$
(Equation~\ref{EqnEta}) is shown as a function of normalised radius $R/R_{\rm e}$
for S\'ersic profiles having $n$=0.5, 1, 2, 3,... 10. 
}
\label{figPet1}
\end{center}
\end{figure}

The Petrosian (1976, his Equation 7) function $\eta(R)$ is given as 
\begin{eqnarray}
\eta(R) = \frac{2\pi \int_{0}^{R} I(R^{\prime}) R^{\prime} {\rm d}R^{\prime} }
{\pi R^2 I(R) }, \\
        = \frac{L(<R)}{\pi R^2 I(R) } = \frac{\langle I \rangle_R}{I(R)}. 
\end{eqnarray} 
It is the average intensity within some projected radius $R$ divided
by the intensity at that radius.  The logarithmic expression is
written as
\begin{equation}
2.5 \log [\eta(R)]=\mu(R) - \langle\mu\rangle_R, 
\label{EqnEta}
\end{equation}
and is shown in Figure~(\ref{figPet1}) for a range of profile shapes $n$.

This is a particular clever quantity because if every galaxy had the
same stellar distribution, such as an $R^{1/4}$ profile, then a radius
where the $\eta$--function equalled some pre--defined, constant value
would correspond to the same number of $R_{\rm e}$ for every galaxy.
Moreover, such measurements are unaffected by such things as
exposure--depth, galactic dust, and cosmological redshift dimming
because they affect both surface brightness terms in Equation~(\ref{EqnEta})
equally.  Even though it is possible to measure the Petrosian radius
without ever assuming or specifying an underlying light--profile
model, the actual form of the stellar distribution is implicitly
incorporated into the Petrosian function and so cannot be ignored 
(as Figure~\ref{figPet1} reveals). 

It turns out the Petrosian function is equal to 
\begin{equation}
  \eta(R) = 2/\alpha(R), 
\end{equation}
where $\alpha(R)$ is given in Equation~(\ref{EqnAlp}; 
Djorgovski \& Spinrad 1981; Djorgovski, Spinrad \& Marr 1984;
Sandage \& Perelmuter 1990, their Section IIa; Kj\ae rgaard, Jorgensen, \& Moles 1993). 
Thus
\begin{equation}
\eta(x,n) = \frac{2n\gamma (2n,x)}{{\rm e}^{-x}x^{2n}}.
\end{equation}

\begin{figure}[ht]
\begin{center}
\includegraphics[scale=0.4, angle=270]{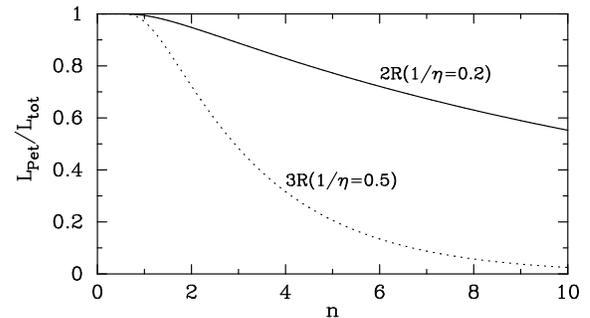}
\caption{
Flux ratio, as a function of light--profile shape $n$, between the
total luminosity $L_{\rm tot}$ 
% obtained by integrating the $R^{1/n}$ profile to infinity 
and the Petrosian luminosity $L_{\rm Pet}$ inside (i) twice the radius
$R_{\rm P}$ where $1/\eta(R_{\rm P})=0.2$ (solid curve) and (ii)
thrice the radius $R_{\rm P}$ where $1/\eta(R_{\rm P})=0.5$ (dotted
curve).
}
\label{figPetL}
\end{center}
\end{figure}

The flux within twice the radius $R_{\rm P}$ when $1/\eta(R_{\rm
P})=0.2$ is often used to estimate an object's flux (e.g., Bershady,
Jangren, \& Conselice 2000; Blanton et al.\ 2001), as is the flux
within 3$R_{\rm P}$ when $1/\eta(R_{\rm P})=0.5$ (e.g., Conselice,
Gallagher, \& Wyse 2002; Conselice et al.\ 2003).  How well this works
of course depends on the shape of the light--profile, and
Figure~(\ref{figPetL}) shows these approximations to the total
luminosity as a function of the S\'ersic index $n$.  In the case of
2$R_{\rm P}$ when $1/\eta(R_{\rm P})=0.2$, one can see that profiles
with $n$=10 will have their luminosities under-estimated by 44.7\% and
those with $n$=4 by only 17.1\%.  The situation is considerably worse
when using 3$R_{\rm P}$ and $1/\eta(R_{\rm P})=0.5$.  A prescription
to correct for the missing light, beyond one's chosen aperture, 
is detailed in Graham et al.\ (2005).

\subsection{Kron magnitudes} 

Kron (1980) presented the following luminosity--weighted radius, $R_1$, 
which defines the `first moment' of an image 
% (see also Koo 1986, his section III)
\begin{equation}
R_1(R) = \frac{2\pi \int_{0}^{R} I(x) x^2 {\rm d}x }
              {2\pi \int_{0}^{R} I(x) x   {\rm d}x }. 
\label{EqKron1}
\end{equation}
He argued that an aperture of radius twice $R_1$, when $R_1$
is obtained by integrating to a radius $R$ that is 1\% of the sky
flux, contains more than $\sim$90\% of an object's total light, making
it a useful tool for estimating an object's flux.

It is worth noting that considerable confusion exists in the
literature in regard to the definition of $R_1$. To help avoid
ambiguity, we point out that $g(x)$ in Kron's (1980) original equation
refers to $xI(x)$, where $x$ is the radius and $I(x)$ the intensity
profile.  Infante (1987) followed this notation, but confusingly a
typo appears immediately after his Equation 3 where he has written
$g(x)\sim I(x)$ instead of $g(x)\sim xI(x)$.  Furthermore, Equation 3
of Bertin \& Arnouts (1996) is given as $R_1 = \sum RI(R)
/ \sum I(R)$ where the summation is over a two--dimensional aperture
rather than a one--dimensional light--profile.  In the latter case, 
one would have $R_1 = \sum R^2I(R) / \sum RI(R)$.

Using a S\'ersic intensity profile, and substituting in 
$x=b(R/R_{\rm e})^{1/n}$, the numerator can be written as
\begin{eqnarray}
2\pi nI_{\rm e}{\rm e}^bR^3_{\rm e} \gamma(3n,x)/b^{3n}. \nonumber 
\end{eqnarray}
Using Equation~(\ref{Lum}) for the denominator, which is simply the 
enclosed luminosity, 
Equation~(\ref{EqKron1}) simplifies to
\begin{equation}
R_1(x,n) = \frac{R_{\rm e}}{b^n} \frac{\gamma(3n,x)}{\gamma(2n,x)}. 
\label{EqKron2}
\end{equation}

The use of `Kron radii' to determine `Kron magnitudes' has proved very
popular, and SExtractor (Bertin \& Arnouts 1996) obtains its
magnitudes using apertures that are 2.5 times $R_1$.  Recently,
however, it has been reported that such an approach may, in some
instances, be missing up to half of a galaxy's light (Bernstein,
Freedman, \& Madore 2002; Benitez et al.\  2004).
If the total light is understood to be that coming from the
integration to infinity of a light--profile, then what is important is
not the sky-level or isophotal-level one reaches, but the number of
effective radii that have been sampled.

\begin{figure}[ht]
\begin{center}
\includegraphics[scale=0.4, angle=270]{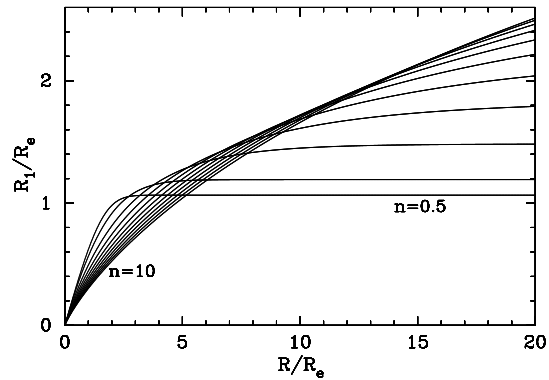}
\caption{
Kron radii $R_1$, as obtained from Equation~(\ref{EqKron2}), are
shown as a function of the radius $R$ to which the integration
was performed.  Values of $n$ range from 0.5, 1, 2, 3,... 10.
}
\label{figKronR}
\end{center}
\end{figure}

For a range of light--profiles shapes $n$, Figure~(\ref{figKronR})
shows the value of $R_1$ (in units of $R_{\rm e}$) as a function of
the number of effective radii to which Equation~(\ref{EqKron1}) has
been integrated.  Given that one usually only measures a
light--profile out to 3--4 $R_{\rm e}$ at best, one can see that only
for light--profiles with $n$ less than about 1 will one come close to
the asymptotic value of $R_1$ (i.e, the value obtained if the profile
was integrated to infinity).  Table~(\ref{Tab1}) shows these 
asymptotic 
values of
$R_1$ as a function of $n$, and the magnitude enclosed within 2$R_1$
and 2.5$R_1$.  This is, however, largely academic because
observationally derived values of $R_1$ will be smaller than
those given in Table~(\ref{Tab1}), at least for
light--profiles with values of $n$ greater than about 1.

\begin{table}[ht]
\begin{center}
\caption{Theoretical Kron Radii and Magnitudes\label{Tab1}}
\begin{tabular}{cccc}
\hline 
S\'ersic $n$  &  $R_1$      &  $L(<2R_1)$  &  $L(<2.5R_1)$ \\
              & [$R_{\rm e}$] &     \%       &     \%        \\
\hline 
         0.5  &  1.06       &     95.7     &    99.3   \\
         1.0  &  1.19       &     90.8     &    96.0   \\
         2.0  &  1.48       &     87.5     &    92.2   \\
         3.0  &  1.84       &     86.9     &    90.8   \\
         4.0  &  2.29       &     87.0     &    90.4   \\
         5.0  &  2.84       &     87.5     &    90.5   \\
         6.0  &  3.53       &     88.1     &    90.7   \\
         7.0  &  4.38       &     88.7     &    91.0   \\
         8.0  &  5.44       &     89.3     &    91.4   \\
         9.0  &  6.76       &     90.0     &    91.9   \\
        10.0  &  8.39       &     90.6     &    92.3   \\
\hline
\end{tabular}
\end{center}
\end{table}

\begin{figure}[ht]
\begin{center}
\includegraphics[scale=0.4, angle=270]{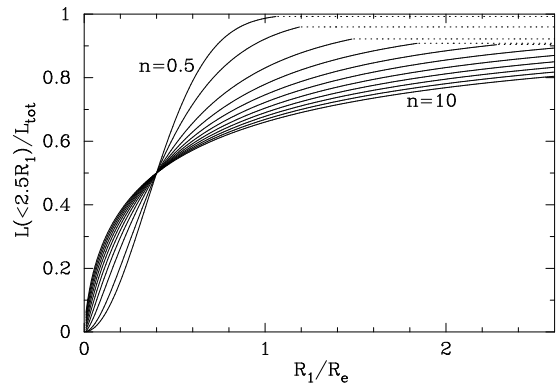}
\caption{
Kron luminosity within 2.5$R_1$, normalised against 
the total luminosity, as a function of how many effective radii $R_1$ 
corresponds to.  Values of $n$ range from 0.5, 1, 2, 3,... 10.
}
\label{figKronL1}
\end{center}
\end{figure}

From Figure~(\ref{figKronR}) one can see, for example, that an
$R^{1/4}$ profile integrated to 4$R_{\rm e}$ will have
$R_1$=1.09$R_{\rm e}$ rather than the asymptotic value of 2.29$R_{\rm
e}$.  Now $2.5 \times 1.09R_{\rm e}$ encloses 76.6\% of the object's
light rather than 90.4\% (see Table~\ref{Tab1}).  This is illustrated
in Figure~(\ref{figKronL1}) where one can see when and how Kron
magnitudes fail to represent the total light of an object.  This
short--coming is worse when dealing with shallow images and with highly
concentrated systems having large values of $n$ (brightest cluster
galaxies are known to have S\'ersic indices around 10 or greater, 
Graham et al.\ 1996). 

\begin{figure}[ht]
\begin{center}
\includegraphics[scale=0.4, angle=270]{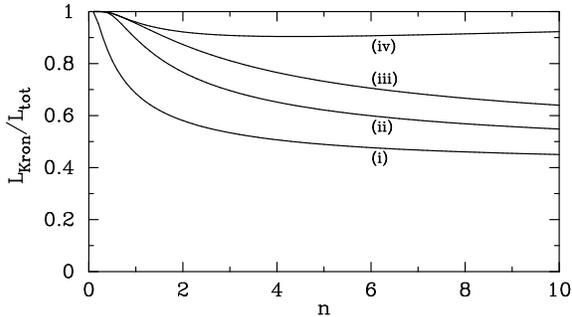}
\caption{
Kron luminosity within 2.5$R_1$, normalised against the total
luminosity, as a function of the underlying light--profile shape $n$.
The different curves arise from the different values of $R1$ obtained
by integrating Equation~(\ref{EqKron1}) to (i) 1$R_{\rm e}$, (ii)
2$R_{\rm e}$, (iii) 4$R_{\rm e}$, and (iv) infinity.
}
\label{figKronL2}
\end{center}
\end{figure}

To provide a better idea of the flux fraction represented by Kron
magnitudes, and one which is more comparable with Figure~(\ref{figPetL}), 
Figure~(\ref{figKronL2}) shows this fraction as a function
of light--profile shape $n$.  The different curves result from
integrating Equation~(\ref{EqKron1}) to different numbers of effective
radii in order to obtain $R1$.  If $n$=4, for example, but one only
integrates out to 1$R_{\rm e}$ (where $R_{\rm e}$ is again understood
to be the true, intrinsic value rather than the observed value), then the
value of $R_1$ is 0.41$R_{\rm e}$ and the enclosed flux within
2.5$R_1$ is only 50.7\%.  If an $n$=10 profile is integrated to only
1$R_{\rm e}$, then $R_1$=0.30$R_{\rm e}$ and the enclosed flux is only
45.0\% within 2.5$R_1$.  It is therefore easy to understand why people
have reported Kron magnitudes as having missed half of an object's
light.

\subsection{The core--S\'ersic model \label{SecCore}}

Due to the presence of partially depleted cores in luminous
($M_B<-20.5$ mag, $H_{\rm 0}=70$ km s$^{-1}$ Mpc$^{-1}$) elliptical
galaxies\footnote{The luminous `core galaxies' likely correspond to
the `bright' family of galaxies identified in Capaccioli, Caon, \&
D'Onofrio et al.\ (1992) and Caon et al.\ (1993).  They tend to have
boxy rather than disky isophotes (Nieto et al.\ 1991), and S\'ersic
indices greater than $\sim$4.  They are commonly understood to be the
product of (elliptical) galaxy mergers (e.g., Capaccioli, Caon, \& D'Onofrio 
1992, 1994; Graham 2004)}, a `core--S\'ersic' model (Graham et al.\ 2003a,b) has been
developed in order to describe and connect the nuclear (typically less
than a few hundred parsecs) and the remaining outer stellar
distribution.  This model consists of an inner power--law and an outer
S\'ersic function, and has proven essential for modeling the {\sl
HST}--resolved light--profiles of luminous early--type galaxies
(Trujillo et al.\ 2004).  As suggested in Graham et al.\ (2003b), the
lower--luminosity `power--law' galaxies have been shown to be
described by the S\'ersic model over their entire radial extent
(Trujillo et al.\ 2004).

\begin{figure}[ht]
\begin{center}
\includegraphics[scale=0.5, angle=270]{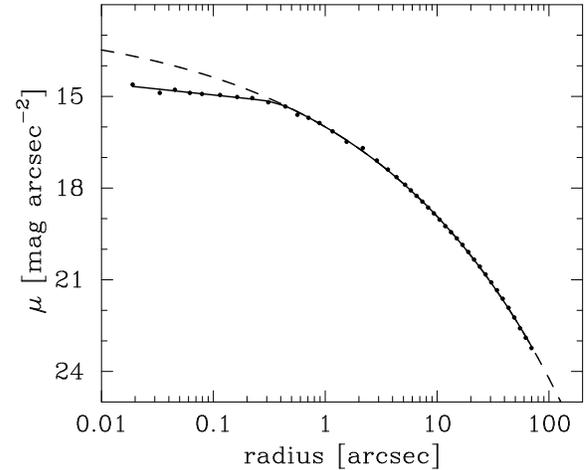}
\caption{
Major--axis, $R$-band light--profile of
NGC~3348.  The solid line is the best--fitting core--S\'ersic model
while the dashed line is the best--fitting S\'ersic model to the data
beyond the {\sl HST}--resolved break radius (Graham et al.\ 2003a,b;
Trujillo et al.\ 2004).
}
\label{fig3348}
\end{center}
\end{figure}

Although the reader is referred to the above papers, especially
the Appendix of Trujillo et al.\ (2004), the core--S\'ersic model 
is given as\footnote{The $\alpha$ and $\gamma$ terms shown here 
should not be confused with those given earlier in the paper, they 
are different quantities.}: 
\begin{equation}
I(R)=I' \left[ 1+\left( \frac{R_b}{R}\right)^{\alpha} \right]^{\gamma/\alpha}
\exp \left\{ -b[(R^{\alpha}+R_b^{\alpha})/R_{\rm e}^{\alpha}]^{1/(\alpha n)}\right\},
\label{bomba}
\end{equation}
where $R_b$ is the break--radius separating the inner power--law having
logarithmic slope $\gamma$ from the outer S\'ersic function.
The intensity $I_b$ at the break--radius $R_b$ can be evaluated
from the expression
\begin{equation}
I' = I_b 2^{-(\gamma/\alpha)} \exp \left[ b(2^{1/\alpha}R_b/R_{\rm e})^{1/n}\right].
\label{iprime}
\end{equation}
The final parameter, $\alpha$, controls the sharpness of the
transition between the inner (power--law) and outer (S\'ersic) regimes
--- higher values of $\alpha$ indicating sharper transitions.  In
practice (e.g., Figure~\ref{fig3348}) we find that a sharp 
transition is adequate and recommend setting 
$\alpha$ to a suitably large constant value (typically anything
greater than 3 is fine), leaving one with a 5--parameter core--S\'ersic
model.

\subsection{Deprojected quantities and dynamical terms}

Ciotti (1991) provides an exact, numerical expression for the
deprojected light--profile of the $R^{1/n}$ model, that is, the
luminosity density profile.  He additionally provides numerical
expressions for the gravitational potential and also the spatial and
line--of--sight velocity dispersions.  These must however be solved by
integration, which means they require considerably more computer time
than analytical expressions.  Ciotti does however provide analytical
expressions for the behavior of the above expressions at both small
and large radii.  Luminosity--weighted aperture velocity dispersions
have been used in Ciotti, Lanzoni, \& Renzini (1996), and also in
Graham \& Colless (1997) where the radial profiles are shown for
different values of the S\'ersic index $n$.
Ciotti (1991) additionally provides expressions for the distribution
function and the normalised differential energy distribution.

An exact, analytical expression for the density profile was finally
discovered a couple of years ago and is given in Mazure \& Capelato
(2002).  It involves the use of somewhat complicated Meijer G
functions.  For those interested in a more simple, analytical
approximation, an accurate expression is given in Prugniel \& Simien
(1997), which is developed slightly in Lima Nieto, Gerbal, \&
M\'arquez (1999) and Trujillo et al.\ (2002).

Mazure \& Capelato (2002) also provide exact analytical expressions
for the mass, gravitational potential, total energy, and the central
velocity dispersion.  For modellers interested in fast--to--compute,
analytical approximations for not only the density profile but also
the potential and force, such expressions, which additionally include
optional power--law cores, can be found elsewhere (B.\ Terzi\'c \& A.W.\
Graham, in preparation).

\section{S\'ersic magnitudes} 

In this article we have compiled and developed equations pertaining to
the S\'ersic profile in a purely analytical manner. To mention just
one of many important uses of the S\'ersic profile is its
potential for deriving accurate total magnitudes.  This need is
motivated by a growing community--wide awareness of the complex nature
of galaxy photometry, and in particular the large amounts of flux
which can be missed by isophotal, aperture, Petrosian, or Kron
magnitudes (e.g., Figures~\ref{figPetL} \& \ref{figKronL2}).  Cross et
al.\ (2004) recently compared APM, SuperCosmos, SDSS and MGC
photometry for several thousand galaxies and concluded that the
photometric errors are mainly dominated by the systematics of the
methodology rather than the random errors.  The S\'ersic magnitude 
provides a logical standard and is
derived by evaluating Equation~\ref{Eqmag} at $R=\infty$ given
$\mu_{\rm e}, n$, and $R_{\rm e}$ derived from a fit to the available
light--profile.

In practice various ``facts-of-life'' issues remain; these are not
specific problems to the S\'ersic model, but generic to studies of
galaxy photometry.  The most obvious ones are: the smoothing effect of
the point-spread function (PSF); profile truncation; multiple
component systems; dust and asymmetric profiles.  All of these can act
to exacerbate or ameliorate the amount of missing flux.  While a
detailed discussion of these issues is beyond the scope of this paper
we provide some suitable references below.

The smearing effect of the PSF will cause the observed profile to tend
to $n \approx 0.5$ (i.e., Gaussian), this is dealt with
straightforwardly by incorporating PSF convolution into the model
fitting process, e.g., Andredakis et al.\ (1995; their Equation 10)
and Trujillo et al.\ (2001a,b). 
Disk truncation is trickier (see the reviews by van der Kruit (2001)
and Pohlen et al.\ 2004) and is assumed to be related to the minimum
gas density required for star-formation (Kennicutt 1989; Kregel \& van
der Kruit 2004).  Initially truncation was reported to occur at around
4 scale--lengths for exponential disks (van der Kruit 1979; van der
Kruit \& Searle 1981).  More recent studies have argued that the
truncation is better described by a broken exponential fit (e.g., de
Grijs et al.\ 2001; Pohlen et al.\ 2002).  Others argue that
truncation may actually be a manifestation of poor background
modelling or simply due to intrinsic variations in the disk (Narayan
\& Jog 2003).  Certainly some recent studies find no discernible
truncation to extremely faint isophotal limits; for example, NGC~300
is a pure exponential out to 10 scale--lengths (Bland-Hawthorn et al.\
2005).  The net result is that in practice it is not clear exactly how
far out one should integrate the S\'ersic profile for disk galaxies.
As discussed in Section~\ref{SecCon}, there is no evidence for
truncation in the elliptical galaxy population.  From
Figure~\ref{figSersic} we see that this issue is far more significant
for high S\'ersic index systems.
A new generation of publicly available 2D fitting codes, in particular
GIM2D
% (fully automated but a restricted profile set; 
(Marleau \& Simard 1998), GALFIT 
% (less automated but a larger profile set
(Peng et al.\ 2002), and BUDDA (de Souza, Gadotti, \& dos Anjos 2004), 
can routinely deal with multiple--component profiles. 
Dust opacity (Disney, Davies, \& Phillipps 1989, Davies et al.\ 1993)
can also lead to changes in the light--profile because of the more
centrally concentrated dust distribution.  Modelling opacity is
non--trivial as there are strong inclination dependencies (Cho\l
oniewski 1991; Jones, Davies \& Trewhella 1996; Graham 2001b) however
models are being developed to provide detailed corrections (e.g.,
Pierini et al.\ 2004; Tuffs et al.\ 2004).  From the dust attenuation
studies of Calzetti (2001, and references therein) and many others,
dust issues can be minimised via structural analysis at near-IR
wavelengths.
Non--biaxial asymmetry of galaxy images can be readily identified via
the use of the `asymmetry' measures (e.g., Schade et al.\ 1995; Rix \&
Zaritsky 1995; Conselice 1997; Kornreich, Haynes, \& Lovelace 1998).

% % It is preferable to embed your figures in the text as in the following example
% \begin{figure}[h]
% \begin{center}
% %\includegraphics[scale=1, angle=0]{figure.eps}
% \caption{An example figure caption.}\label{figexample}
% \end{center}
% \end{figure}

%%Format tables as in the following example
% \begin{table}[h]
% \begin{center}
% \caption{Example Table}\label{tableexample}
% \begin{tabular}{lcc}
% \hline Column 1 & Column 2 & Column 3 \\
% \hline Table Content$^a$ \\
% \hline
% \end{tabular}
% \medskip\\
% $^a$Table footnotes go here.\\
% \end{center}
% \end{table}

\section*{Acknowledgments} 
We kindly thank Valeria Coenda for faxing us a copy of S\'ersic's
1963 article, referenced in his 1968 Atlas. 
We are additionally grateful to Steve Phillipps for refereeing this work, 
and Massimo Capaccioli for his helpful and 
rapid response to our request for information and references.

%\end{multicols}

\end{document}